# Dynamic phase microscopy: measurements of translational displacements at sub-nanometer scale

**V.P.Tichinsky and A.V. Kretushev**

*Moscow State Institute for Radioengineering, Electronics and Automation, prosp. Vernadskogo 78, 119454 Moscow, Russia*
*vtych@yandex.ru*

**P.N. Luskinovich**

*A.M. Prokhorov Institute of General Physics of Russian Academy of Sciences,
ul. Vavilova 38, 119991 Moscow, Russia*

**Abstract:** Dynamic phase microscopy has been applied for measurements of nanometer-scale displacements of a piezoelectric scanner. This scanner, which was designed for calibration purposes for scanning probe microscopy and TEM, exhibited a linear and hysteresis-free translation in the 0.05 – 20 nm range. The voltage-activated motion is described by a coefficient of 0.03±0.005 nm/V.

© 2006 Optical Society of America

___________________________________________________________________________________________

___________________________________________________________________________________________

## 1. Introduction

Developments in nanotechnology are largely relying on characterization techniques such as TEM and scanning probe microscopy, which enable measurements at the nanometer scale. Atomic force microscopy (AFM) is the leading scanning probe technique and it is widely applied for real time visualization of nanostructures and various dynamic processes at small scales [1]. Piezoelectric actuators, which are used in AFM, are made of polycrystalline materials and possess a number of imperfections (hysteresis, non-linearity, etc) limiting accuracy of their translation [2]. At present, open loop scanners of scanning probe microscopes are calibrated at scales above 100 nm with different gratings and at the atomic-scale with crystalline samples like mica (spacing 0.52 nm) or graphite (spacing of 0.25 nm).

The calibration data are used for software that corrects the actual motion of the scanners during imaging. In close loop scanners their actual displacement is measured independently and used for correction of the scanner motion. Yet the close loop scanners are noisier than the open loop ones and less useful for imaging of surfaces with resolution down to 1 nm.

Therefore, high-resolution AFM imaging is performed with the open loop scanners that require calibration means in the 1 – 100 nm range.

One of the possible solutions for the calibration of AFM scanners and TEM microscopes is the use of linear hysteresis-free piezoelectric actuators based on single crystals [3]. The control of the motion of such actuators can be performed with dynamic phase microscopy method and this procedure is described in the paper. Once calibrated this actuator can be a useful tool in microscopy labs. It also can be considered as a scanner in two-scanner systems, where imaging of high-resolution is performed with a small scan actuator and visualization of larger structures with an actuator based on polycrystalline material.

There is another important issue to address is providing a nanotechnology science with metrological devices, namely static dimensions standards and dynamic standards for lateral displacement measurements. The conventional static standards like diffraction lattices have the hindrances either in manufacturing process of the regular structures, and in keeping of the parameters of these structures safe out of vacuum and cleaning rooms. At ambient conditions, even the finest filters being applied the surface is covered with adsorbed layer or with the tiny dust particles of submicron size as well, which capable to destroy measurement accuracy. To clean a specimen surface it is necessary to use special technological processes, for instance ion purification, which in turn can lead to surface deformation.

The implementation of dynamic standards [4] is a unique possibility to solve the issues mentioned above. In this case a displacement of a surface investigated is carried out by driven voltage in an ambient equilibrium conditions without additional safety expenses. Moreover, in contrast to regular measure devices based on silicon surface's structures the dynamic etalons capable to provide translational movements not only at nanometer but at picometer scale as well. However, a real-time registration and precise measurement of such small-scale translations by optical methods are not possible for Legal Metrology Services by now. Yet, it becomes possible due to unique properties of phase imaging and multiple super resolution [5] when dynamic phase microscopy (DPM) method [6-8] is applied.

For the first time the certifications results of dynamic standard "Nanotester" [4] at sub-nanometer scale by new nontraditional method are presented in this article. The purposes of the "Nanotester" standard measurements given below are to highlight the capabilities of new dynamic phase microscopy method, hysteresis absence evidence, and to show the transmission constant measurement procedure. The fundamental application fields of "Nanotester" due to high accuracy of measurements at sub-nanometer scale are broad: scanning probe and electron microscopy, optical interferometry, nanotechnology, nanoelectronics, nanooptics, nanomechanics. It can be successfully applied for testing the dynamic characteristics of the probe stabilizing systems and interferometers.

**2. Nanotester and its calibration**

"Nanotester" cell (Fig. 1) is designed to generate vertical or horizontal displacements in the 0-20 nm and 0-60 nm ranges, respectively. These are the motions of a surface of a central block carrying a piezoelectric crystal in the horizontal and vertical cells. An electronic block (not shown) supplied voltage ($\pm 1$ kV) to the cells to cause the crystal motion. These cells are calibrated with a coherent phase microscope "Airyscan" [7, 8] with the helium - neon laser ($\lambda = 632,8$ nm) as a source in a setup described in Fig. 2. An optical-path difference was measured using modulation method with position-sensible photodetector – dissector image tube. In static images the microscope provided sensitivity about 0.5 nm and a spatial resolution up to 100 nm. Sampling rate was limited with modulation frequency 1 kHz (or 1 ms/pixel). The software allowed carrying out the statistical analysis of space-time processes.

This instrument is used for dynamic phase measurements in which 3D surface pattern is represented by a function $h(X,Y,t)$, where $h$ – optical path difference measured by the interferometer and $X,Y$ – surface coordinates. Most of the calibration operations for displacements in the nanometer range require low electric and vibration noise environment and high temperature stability eliminating thermal drift.

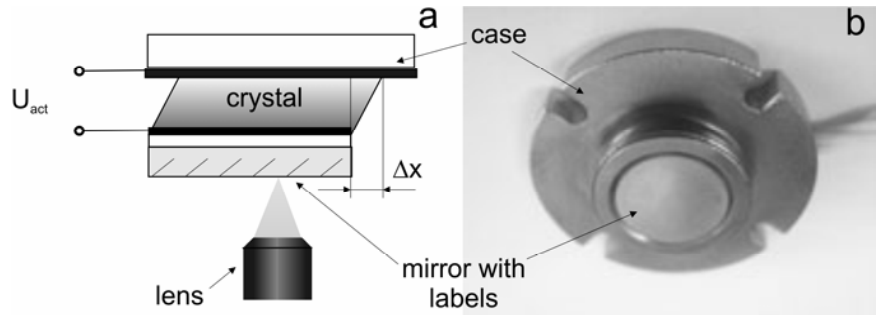

Fig. 1. Design and photo of dynamic "Nanotester" standard.

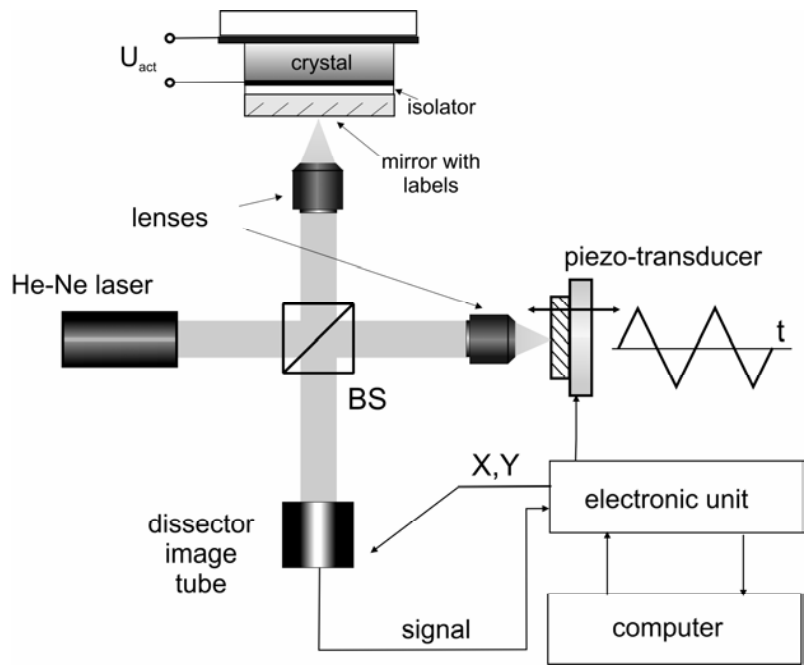

Fig. 2. Setup of Coherent Phase Microscope "Airyscan". Measurements of an optical-path difference are carried out by compensational method in each point consistently scanning images. Optical-path difference values normalized on the wavelength are collected in computer memory.

### 3. Displacements in the 1 nm - 60 nm range

Small lateral motions of the crystal of "Nanotester" are registered as a shift of a surface reference feature. It can be also done by measurements of vertical displacement of a structure with height gradient. In this case a lateral motion will cause a known change of height at the reference location. The structures of steep slopes are desirable for more accurate measurements.

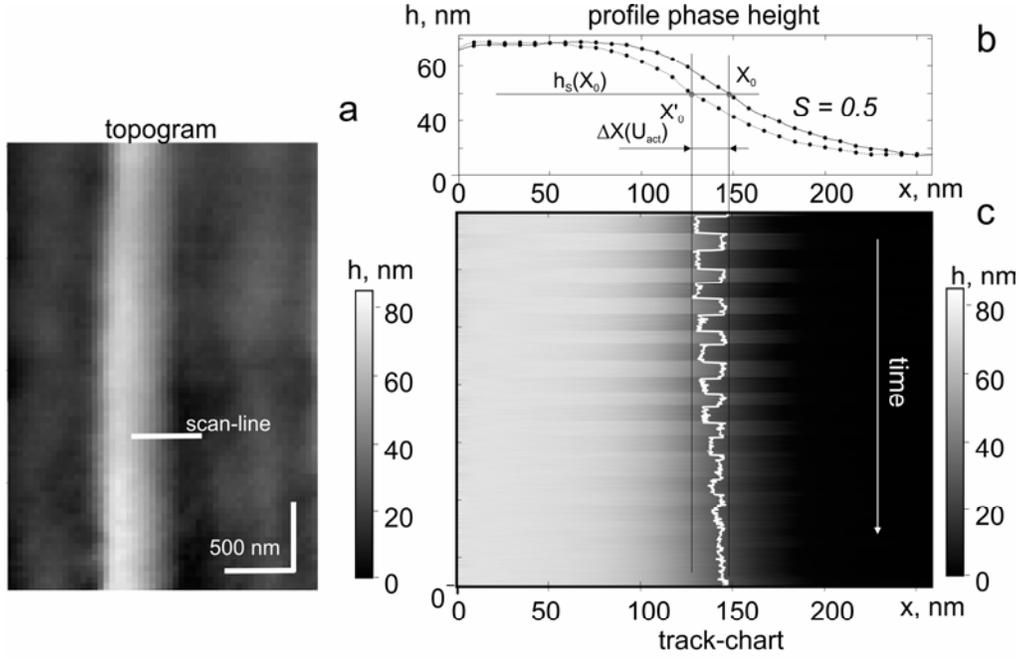

Fig. 3. Measurements of lateral translations. a - 2D image of "Nanotester" standard surface with micro defect and scan-line Y – const along which the phase height profile were periodically (30 ms) measured. b - Two shifted phase height of micro defect. c - track-chart (sequence of the phase height profiles) with periodic commutations of the voltage with decreasing amplitude. The trace of the track-chart section by horizontal plane (light contour line) shows alteration of shift in course of voltage manipulation.

A 2D phase image of a defect is shown in Fig. 3a and "scan-line" $Y = const$ position, along which phase height profiles $h(X,Y)$ and $h(X+\Delta X,Y)$ and profile difference $\Delta h(X)$ were defined in image section. Fig. 3b shows two profiles with gradient $S = 0.5$, where their shift $\Delta X$ from the voltage is noticeable. Alterations of local height $h(X,t)$ as the applied voltage changed from 545V to 51V is demonstrated by a series of profiles. Light contour line (see Fig. 3c) shows lateral time shift of the point position fixed on the level of maximum gradient of the profile. Minimal quasi-static lateral shifts measured by these methods were limited by their intrinsic noise level and were of 1nm.

**4. Displacements below 1 nm**

Displacements in pico-meter range were recorded with a modulation technique. The response of the piezoelectric crystal to sine wave voltage with fixed amplitude $U$ and frequency $f$ was collected during 3 minutes. Fourier analysis of the response at fixed phase level provides spectral density distribution of the shift $\rho(f,U)$ and frequency $f = 1\ Hz$. (Linear interpolation has been used for a reduction of a digitization step to 0.1 nm.) The displacement $\Delta X$ was determined from the following equation:

$$X(U) = \sqrt{\frac{1}{T} \cdot \int_{f_1}^{f_2} \frac{\rho(f,U)}{f} df}, \qquad (1)$$

where $\rho(f,U)$ spectral density, $T$ – registration time, $f_1$ and $f_2$ – the lowest and highest frequency of the signal spectrum.

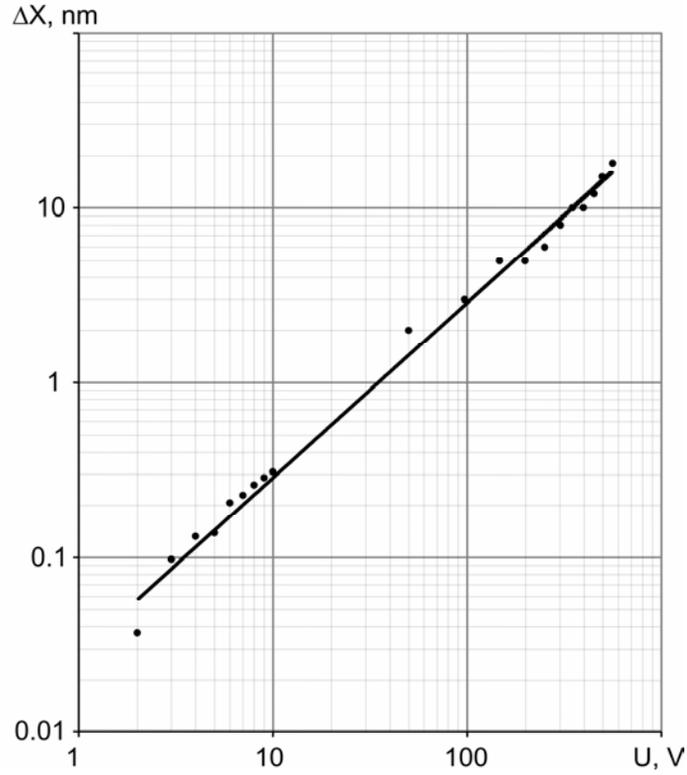

Fig. 4. The results of standard "Nanotester" certification. Measurements of sub nanometer displacements $\Delta X(U)$ were done by modulation method with periodic (1 Hz) changes of the voltage.

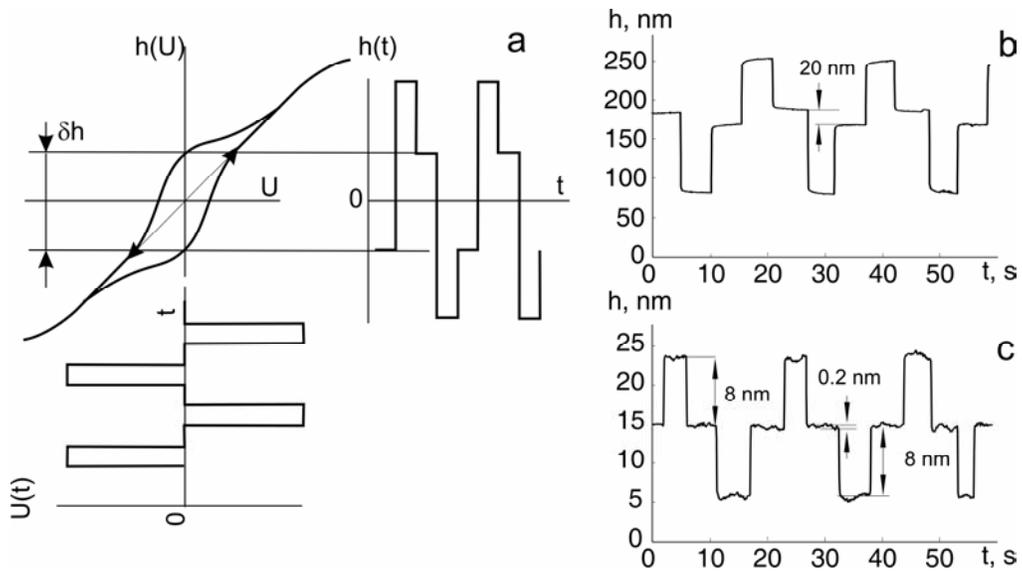

Fig. 5. Measurements of hysteresis. a - typical for piezoceramics hysteresis loop is cause of ambiguity in $\Delta h(U)$ positioning. b - measurements of hysteresis of an ordinary piezoceramic transducer verified adequateness of the method. c- the measurements show the hysteresis-free translations of "Nanotester" standard.

Fig. 4 shows the results of the etalon certification by both methods, consequently, minimum (approx: 0.06 nm) range of lateral shifts registered by modulation method was four orders lower than classic lateral resolution of the lens (Airy disk radius) – 400 nm.

**Evaluation of crystal hysteresis**

For evaluation of a possible hysteresis of the piezoelectric crystal we measured small differences $\delta h = \Delta h(U = + 0\ V) - \Delta h(U = - 0\ V)$ of the crystal surface position at zero voltage after applying voltage of opposite sign. Functional dependence $h(U)$ with hysteresis loop typical for piezoceramic is sketched out on Fig. 5a. Low values $\delta h$ measurements require low noise level, high accuracy and stability of the instruments.

The next step was to verify the above method using an ordinary piezoceramic transducer. The results measurement of axial movements as height alterations $h(t)$ with fixation of zero values are shown on Fig. 5b. The hysteresis (the difference of zero values) amounts to 20 nm. The similar measurements on the "Nanotester" (see Fig. 5c) indicated practical absence of hysteresis $(\delta X < 0.2\ nm)$.

**Acknowledgments**

The authors would like to thank Dr. Magonov S.N. for valuable corrections and "SIA Technosystems LV" for the provision of "Nanotester" standards. This work was partly supported by Russian Foundation for Basic Researches (Grant # 04-04-49132)